\documentstyle[aps,epsf]{revtex}
\input{psfig}
\def\gsim{\mathrel{\raise.3ex\hbox{$>$}\mkern-14mu
             \lower0.6ex\hbox{$\sim$}}}
\def\lsim{\mathrel{\raise.3ex\hbox{$<$}\mkern-14mu
             \lower0.6ex\hbox{$\sim$}}}

\begin{document}

\title{Quantum Field Theory\thanks{To appear in the American Physical
Society Centenary issue of Reviews of Modern Physics, March 1999.}}

\author{Frank Wilczek\thanks{wilczek@ias.edu~~~IASSNS-HEP 98/20}}
\address{Institute for Advanced Study, School of Natural Science, Olden Lane, 
Princeton, NJ 08540}
\maketitle

\begin{abstract}

I discuss the general principles underlying quantum field theory, and
attempt to identify its most profound consequences.  The deepest of
these consequences result from the infinite number of degrees of
freedom invoked to implement locality.  I mention a few of its most
striking successes, both achieved and prospective.  Possible
limitations of quantum field theory are viewed in the light of its history.
\end{abstract}

\section{Survey}

Quantum field theory is the framework in which the regnant theories of
the electroweak and strong interactions, which together form the
Standard Model, are formulated.  Quantum electrodynamics (QED),
besides providing a complete foundation for atomic physics and
chemistry, has supported calculations of physical quantities with
unparalleled precision.  The experimentally measured value of the magnetic
dipole moment of the muon, 
\begin{equation}
(g_\mu -2)_{\rm exp.~} ~=~
233 ~184 ~600~ (1680) \times 10^{-11},
\end{equation}
for example,
should be compared with the theoretical prediction
\begin{equation}
(g_\mu-2)_{\rm theor.} ~=~ 233~ 183~ 478 ~~(308)
\times 10^{-11}.
\end{equation}

In quantum chromodynamics (QCD) we
cannot, for the forseeable future, aspire to to comparable accuracy.
Yet QCD provides different, and at least equally impressive, evidence
for the validity of the basic principles of quantum field theory.
Indeed, because in QCD the interactions are stronger, QCD
manifests a wider variety of phenomena characteristic of quantum field
theory.  These include especially running of the effective coupling
with distance or energy scale and the phenomenon of confinement.  QCD
has supported, and rewarded with experimental confirmation, both
heroic calculations of multi-loop diagrams and massive numerical
simulations of (a discretized version of) the complete theory.

Quantum field theory also provides powerful tools for condensed matter
physics, especially in connection with the quantum many-body problem
as it arises in the theory of metals, superconductivity, the
low-temperature behavior of the quantum liquids He$^3$ and He$^4$, and
the quantum Hall effect, among others.  Although for reasons of space
and focus I will not attempt to do justice to this aspect here, the
continuing interchange of ideas between condensed matter and high
energy theory, through the medium of quantum field theory, is a
remarkable phenomenon in itself.  A partial list of historically
important examples includes global and local spontaneous symmetry
breaking, the renormalization group, effective field theory, solitons,
instantons, and fractional charge and statistics. 

It is clear, from all
these examples, that quantum field theory occupies a central position
in our description of Nature.  It provides both our best working
description of fundamental physical laws, and a fruitful tool for
investigating the behavior of complex systems.  But the enumeration of
examples, however triumphal, serves more to pose than to answer more
basic questions: What are the essential features of quantum field
theory?  What does quantum field theory add to our understanding of
the world, that was not already present in quantum mechanics and
classical field theory separately?  

The first question has no sharp
answer.  Theoretical physicists are very flexible in adapting their
tools, and no axiomization can keep up with them.  However I think it
is fair to say that the characteristic, core ideas of quantum field
theory are twofold.  First, that the basic dynamical degrees of
freedom are operator functions of space and time -- quantum fields,
obeying appropriate commutation relations.  Second, that the
interactions of these fields are local.  Thus the equations of motion
and commutation relations governing the evolution of a given quantum
field at a given point in space-time should depend only on the
behavior of fields and their derivatives at that point.  One might
find it convenient to use other variables, whose equations are not
local, but in the spirit of quantum field theory there must always be
some underlying fundamental, local variables.  These ideas, combined
with postulates of symmetry (e.g., in the context of the standard
model, Lorentz and gauge invariance) turn out to be amazingly
powerful, as will emerge from our further discussion below.  

The field concept came to dominate physics starting with the work of
Faraday in the mid-nineteenth century.  Its conceptual advantage over
the earlier Newtonian program of physics, to formulate the fundamental
laws in terms of forces among atomic particles, emerges when we take
into account the circumstance, unknown to Newton (or, for that matter,
Faraday) but fundamental in special relativity, that influences travel
no farther than a finite limiting speed.  For then the force on a
given particle at a given time cannot be deduced from the positions of
other particles at that time, but must be deduced in a complicated way
from their previous positions.  Faraday's intuition that the
fundamental laws of electromagnetism could be expressed most simply in
terms of fields filling space and time was of course brilliantly
vindicated by Maxwell's mathematical theory.

The concept of locality, in the crude form that one can predict the
behavior of nearby objects without reference to distant ones, is basic
to scientific practice.  Practical experimenters -- if not astrologers
-- confidently expect, on the basis of much successful experience,
that after reasonable (generally quite modest) precautions to isolate
their experiments they will obtain reproducible results.  Direct
quantitative tests of locality, or rather of its close cousin
causality, are afforded by dispersion relations.

The deep and ancient historic roots of the field and locality concepts
provide no guarantee that these concepts remain relevant or valid when
extrapolated far beyond their origins in experience, into the
subatomic and quantum domain.  This extrapolation must be judged by
its fruits.  That brings us, naturally, to our second question.

Undoubtedly the single most profound fact about Nature
that quantum field theory uniquely explains is {\it the existence of
different, yet indistinguishable, copies of elementary particles}.
Two electrons anywhere in the Universe, whatever their origin or
history, are observed to have exactly the same properties.  We
understand this as a consequence of the fact that both are excitations
of the same underlying ur-stuff, the electron field.  The electron
field is thus the primary reality.  The same logic, of course, applies
to photons or quarks, or even to composite objects such as atomic
nuclei, atoms, or molecules.  The indistinguishability of particles is
so familiar, and so fundamental to all of modern physical science,
that we could easily take it for granted.  Yet it is by no means
obvious.  For example, it directly contradicts one of the pillars of
Leibniz' metaphysics, his ``principle of the identity of
indiscernables,'' according to which two objects cannot differ
solely in number.  And Maxwell thought the similarity of different
molecules so remarkable that he devoted the last part of his {\it
Encyclopedia Brittanica\/} entry on Atoms -- well over a thousand
words -- to discussing it.  He concluded that ``the formation of a
molecule is therefore an event not belonging to that order of nature
in which we live ... it must be referred to the epoch, not of the
formation of the earth or the solar system ... but of the
establishment of the existing order of nature ...'' 

The existence of
classes of indistinguishable particles is the necessary logical
prerequisite to a second profound insight from quantum field theory:
{\it the assignment of unique quantum statistics\/} to each class.
Given the indistinguishability of a class of
elementary particles, and complete invariance of their interactions
under interchange, the general principles of quantum mechanics teach
us that solutions forming any representation of the permutation
symmetry group retain that property in time, but do not constrain
which representations are realized.  Quantum field theory not only
explains the existence of indistinguishable particles and the
invariance of their interactions under interchange, but also
constrains the symmetry of the solutions.  For bosons only the
identity representation is physical (symmetric wave functions), for
fermions only the one-dimensional odd representation is physical
(antisymmetric wave functions).  One also has the spin-statistics
theorem, according to which objects with integer spin are bosons,
whereas objects with half odd integer spin are fermions.  Of course,
these general predictions have been verified in many experiments.  The
fermion character of electrons, in particular, underlies the stability
of matter and the structure of the periodic table.  

A third profound
general insight from quantum field theory is {\it the existence of
antiparticles}.  This was first inferred by Dirac on the basis of a
brilliant but obsolete interpretation of his equation for the electron
field, whose elucidation was a crucial step in the formulation of
quantum field theory.   In quantum field theory, we re-interpret the Dirac wave
function as a position (and time) dependent operator.  It can be
expanded in terms of the solutions of the Dirac equation, with
operator coefficients.  The coefficients of positive-energy solutions
are operators that destroy electrons, and the coefficients of the
negative-energy solutions are operators that create positrons (with
positive energy).  With this interpretation, an improved version of
Dirac's hole theory 
emerges in a straightforward way. (Unlike the original hole theory,
it has
a sensible generalization to bosons, and to
processes where the number of electrons minus positrons changes.)
A very general consequence of
quantum field theory, valid in the presence of arbitrarily complicated
interactions, is the CPT theorem.  It states that the product of
charge conjugation, parity, and time reversal is always a symmetry of
the world, although each may be -- and is! -- violated separately.
Antiparticles are strictly defined as the CPT conjugates of their
corresponding particles.  

The three outstanding facts we have
discussed so far: the existence of indistinguishable particles, the
phenomenon of quantum statistics, and the existence of antiparticles,
are all essentially consequences of {\it free\/} quantum field theory.
When one incorporates interactions into quantum field theory, two
additional general features of the world immediately become brightly
illuminated.  

The first of these is {\it the ubiquity of particle
creation and destruction processes}.  Local interactions involve
products of field operators at a point.  When the fields are expanded
into creation and annihilation operators multiplying modes, we see
that these interactions correspond to processes wherein particles can
be created, annihilated, or changed into different kinds of particles.
This possibility arises, of course, in the primeval quantum field
theory, quantum electrodynamics, where the primary interaction arises
from a product of the electron field, its Hermitean conjugate, and the
photon field.  Processes of radiation and absorption of photons by
electrons (or positrons), as well as electron-positron pair creation,
are encoded in this product. Just because the emission and absorption
of light is such a common experience, and electrodynamics such a
special and familiar classical field theory, this correspondence
between formalism and reality did not initially make a big
impression. The first conscious exploitation of the potential for
quantum field theory to describe processes of transformation was
Fermi's theory of beta decay.  He turned the procedure around,
inferring from the observed processes of particle transformation the
nature of the underlying local interaction of fields.  Fermi's
theory involved creation and annihilation not of photons, but of
atomic nuclei and electrons (as well as neutrinos) -- the ingredients
of `matter'.  It began the process whereby classic atomism,
involving stable individual objects, was replaced by a more
sophisticated and accurate picture.  In this picture it is only the
fields, and not the individual objects they create and destroy, that
are permanent.  

The second is {\it the association of forces and
interactions with particle exchange}.  When Maxwell completed the
equations of electrodynamics, he found that they supported source-free
electromagnetic waves.  The classical electric and magnetic fields
thus took on a life of their own.  Electric and magnetic forces
between charged particles are explained as due to one particle acting
as a source for electric and magnetic fields, which then influence
others.  With the correspondence of fields and particles, as it arises
in quantum field theory, Maxwell's discovery corresponds to the
existence of photons, and the generation of forces by intermediary
fields corresponds to the exchange of virtual photons.  The
association of forces (or, more generally, interactions) with exchange
of particles is a general feature of quantum field theory.  It was
used by Yukawa to infer the existence and mass of pions from the range
of nuclear forces, and more recently in electroweak theory to infer
the existence, mass, and properties of W and Z bosons prior to their
observation, and in QCD to infer the existence and properties of gluon
jets prior to their observation.

The two additional outstanding facts we just discussed: the
possibility of particle creation and destruction, and the association
of particles with forces, are essentially consequences of classical
field theory supplemented by the connection between particles and
fields we learn from free field theory.  Indeed, classical waves with
nonlinear interactions will change form, scatter, and radiate, and
these processes exactly mirror the transformation, interaction, and
creation of particles.  In quantum field theory, they are properties
one sees already in {\it tree graphs}.

The foregoing major consequences of free quantum field
theory, and of its formal extension to include nonlinear interactions,
were all well appreciated by the late 1930s.  The deeper properties of
quantum field theory, which will form the subject of the remainder of
this paper, arise from the need to introduce {\it infinitely many
degrees of freedom}, and the possibility that all these degrees of
freedom are excited as quantum-mechanical fluctuations.  From a
mathematical point of view, these deeper properties arise when we
consider {\it loop graphs}.  

>From a physical point of view, the potential pitfalls associated with
the existence of an infinite number of degrees of freedom first showed
up in connection with the problem which led to the birth of quantum
theory, that is the ultraviolet catastrophe of blackbody radiation
theory.  Somewhat ironically, in view of later history, the crucial
role of the quantum theory here was to remove the disastrous
consequences of the infinite number of degrees of freedom possessed by
classical electrodynamics.  The classical electrodynamic field can be
decomposed into independent oscillators with arbitrarily high values
of the wavevector.  According to the equipartition theorem of
classical statistical mechanics, in thermal equilibrium at temperature
$T$ each of these oscillators should have average energy $kT$.
Quantum mechanics alters this situation by insisting that the
oscillators of frequency $\omega$ have energy quantized in units of
$\hbar \omega$.  Then the high-frequency modes are exponentially
suppressed by the Boltzmann factor, and instead of $kT$ receive
${\hbar \omega e^{-{\hbar \omega\over kT}} \over 1- e^{-{\hbar
\omega\over kT}}}$.  The role of the quantum, then, is to prevent
accumulation of energy in the form of very small amplitude excitations
of arbitrarily high frequency modes.  It is very effective in
suppressing the {\it thermal\/} excitation of high-frequency modes.

But while removing arbitrarily small amplitude
excitations, quantum theory introduces the idea that the modes are
always {\it intrinsically\/} excited to a small extent, proportional
to $\hbar$.  This so-called zero point motion is a consequence of the
uncertainty principle.  For a harmonic oscillator of frequency
$\omega$, the ground state energy is not zero, but ${1\over 2} \hbar
\omega$.  In the case of the electromagnetic field this leads, upon
summing over its high-frequency modes, to a highly divergent total
ground state energy.  For most physical purposes the absolute
normalization of energy is unimportant, and so this particular
divergence does not necessarily render the theory useless.  \footnote{One
would think that gravity should care about the absolute normalization
of energy.  The zero-point energy of the electromagnetic field, in
that context, generates an infinite cosmological constant.  This might
be cancelled by similar negative contributions from fermion fields, as
occurs in supersymmetric theories, or it might indicate the need for
some other profound modification of physical theory.}  It does,
however, illustrate the dangerous character of the high-frequency
modes, and its treatment gives a first indication of the leading theme
of renormalization theory: we can only require -- and generally will
only obtain -- sensible, finite answers when we ask questions that
have direct, operational physical meaning.  

The existence of an
infinite number of degrees of freedom was first encountered in the
theory of the electromagnetic field, but it is a general phenomenon,
deeply connected with the requirement of locality in the interactions
of fields.  For in order to construct the local field $\psi (x)$ at a
space-time point $x$, one must take a superposition
\begin{equation}
\psi(x) ~=~ \int {d^4k\over (2\pi )^4} e^{ikx} \tilde
\psi (k)
\end{equation}
that includes field components $\tilde \psi (k)$
extending to arbitrarily large momenta.  Moreover in a generic
interaction
\begin{equation}
\int {\cal L} = \int \psi (x)^3 = \int
{d^4k_1\over (2\pi )^4}{d^4k_2\over (2\pi )^4}{d^4k_3 \over (2\pi )^4}
\tilde \psi (k_1) \tilde \psi (k_2) \tilde \psi (k_3) (2\pi
)^4\delta^4(k_1 + k_2 +k_3)
\end{equation}
we see that a low momentum
mode $k_1 \approx 0 $ will couple without any suppression factor to
high-momentum modes $k_2$ and $k_3 \approx -k_2$.  Local couplings are
``hard'', in this sense. Because locality requires the existence
of infinitely many degrees of freedom at large momenta, with hard
interactions, ultraviolet divergences similar to the ones cured by
Planck, but driven by quantum rather than thermal fluctuations, are
never far off stage.  As mentioned previously, the deeper physical
consequences of quantum field theory arise from this circumstance.

First of all, it is much more difficult to construct non-trivial
examples of interacting relativistic quantum field theories than
purely formal considerations would suggest.  One finds that {\it the
consistent quantum field theories form a quite limited class, whose
extent depends sensitively on the dimension of space-time and the
spins of the particles involved}.  Their construction is quite
delicate, requiring limiting procedures whose logical implementation
leads directly to renormalization theory, the running of couplings,
and asymptotic freedom.  

Secondly, {\it even those quantum theories
that can be constructed display less symmetry than their formal
properties would suggest}.  Violations of naive scaling relations --
that is, ordinary dimensional analysis -- in QCD, and of baryon number
conservation in the standard electroweak model are examples of this
general phenomenon.  The original example, unfortunately too
complicated to explain fully here, involved the decay process $\pi^o
\rightarrow \gamma \gamma$, for which chiral symmetry (treated
classically) predicts much too small a rate.  When the correction
introduced by quantum field theory (the so-called `anomaly') is
retained, excellent agreement with experiment results.   

These deeper consequences of quantum field
theory, which might superficially appear rather technical, largely
dictate the structure and behavior of the Standard Model -- and,
therefore, of the physical world.  My goal in this preliminary survey
has been to emphasize their profound origin.  In the rest of the
article I hope to convey their main implications, in as simple and
direct a fashion as possible.

\section{ Formulation}

The physical constants $\hbar$ and $c$ are so deeply embedded in the
formulation of relativistic quantum field theory that it is standard
practice to declare them to be the units of action and velocity,
respectively.  In these units, of course, $\hbar = c = 1$.   With this
convention,  all physical quantities of interest have units which are
powers of mass.  Thus the dimension of momentum is (mass)$^1$ or
simply 1, since mass$\times c$ is a momentum, and the dimension of
length is  (mass)$^{-1}$ or simply -1, since $\hbar c$/mass is a
length.  The usual way to construct quantum field theories is by
applying the rules of quantization to a continuum field theory,
following the canonical procedure of replacing Poisson brackets by
commutators (or, for fermionic fields, anticommutators).  The field
theories that describe free spin 0 or free spin ${1\over 2}$ fields of
mass $m, \mu$ respectively are based on the Lagrangian
densities
\begin{equation}{\cal L}_0 (x)~=~ {1\over 2} \partial_\alpha
\phi(x) \partial^\alpha \phi(x)  - {m^2\over 2}
\phi(x)^2\label{eq:spinzero}
\end{equation}
\begin{equation}{\cal
L}_{1\over 2}(x) ~=~ \bar \psi(x) (i \gamma^\alpha \partial_\alpha  -
\mu) \psi(x) .\label{eq:spinhalf}
\end{equation}

Since the action $\int
d^4x {\cal L}$ has mass dimension 0, the mass dimension of a scalar
field like $\phi$ is 1 and of a spinor field like $\psi$ is ${3\over
2}$.  For free spin 1 fields the Lagrangian density is that of
Maxwell,
\begin{equation}{\cal L}_1 (x) ~=~ -{1\over 4}
(\partial_\alpha A_\beta (x) - \partial_\beta A_\alpha (x) )
(\partial^\alpha A^\beta (x) - \partial^\beta A^\alpha (x)
),\label{eq:spinone}
\end{equation}
so that the mass dimension of the
vector field $A$ is 1.   The same result is true for non-abelian
vector fields (Yang-Mills fields).  

Thus far all our Lagrangian
densities have been quadratic in the fields. Local interaction terms
are obtained from Lagrangian densities involving products of fields
and their derivatives at a point.   The coefficient of such a term is
a coupling constant, and must have the appropriate mass dimension so
that  the Lagrangian density has mass dimension 4.  Thus the mass
dimension of  a Yukawa coupling $y$, which multiplies the product of
two spinor fields and a scalar field, is zero.   Gauge couplings $g$
arising in the minimal coupling procedure $\partial_\alpha \rightarrow
\partial_\alpha + i g A_\alpha$ are also evidently of mass dimension
zero.    

The possibilities for couplings with non-negative mass
dimension are very restricted. This fact is quite important, for
the following reason.  Consider the effect of treating a given
interaction term as a perturbation.  If the coupling $\kappa$
associated to this interaction has negative mass dimension -p, then
successive powers of it will occur in the form of powers of 
$\kappa \Lambda^p$,
where $\Lambda$ is some parameter with dimensions of mass.  Because,
as we have seen, the interactions in a local field theory are hard, we
can anticipate that $\Lambda$ will characterize the largest mass scale
we allow to occur (the cutoff),   and will diverge to infinity as the
limit on this mass scale is removed.  So we expect that it will be
difficult to make sense of fundamental interactions having negative
mass dimensions, at least in perturbation theory.  Such interactions
are said to be nonrenormalizable.   

The standard model is formulated
entirely using renormalizable interactions.    It has been said that
this is not in itself a fundamental fact about nature.  For if
non-renormalizable interactions occurred in the effective description
of physical behavior below a certain mass scale, it would simply mean
that the theory must change its nature -- presumably by displaying new
degrees of freedom --  at some larger mass scale.   If we adopt this
point of view, the significance of the fact that the standard model
contains only renormalizable operators is that it does not require
modification up to arbitrarily high scales (at least on the grounds of
divergences in perturbation theory).    
Whether or not we call this a fundamental fact, it is certainly a
profound one.  

Moreover, all the renormalizable interactions
consistent with the gauge symmetry and multiplet structure of the
standard model do seem to occur -- ``what is not forbidden, is
mandatory''.   There is a beautiful agreement between the symmetries
of the standard model, allowing arbitrary renormalizable interactions,
and the symmetries of the world.  One understands why strangeness is
violated, but baryon number is not.  (The only discordant element  is
the so-called $\theta$ term of QCD, which is allowed by the symmetries
of the standard model but is measured to be quite accurately zero.   
A plausible solution to this problem exists.  It involves a
characteristic very light {\it axion\/} field.) 

The power counting
rules for estimating divergences assume that there are no special
symmetries cancelling off the contribution of high energy modes.  They
do not apply, without further consideration, to supersymmetric
theories, in which the contributions of boson and fermionic modes
cancels, nor to theories derived from supersymmetric theories by soft
supersymmetry breaking.  In the latter case the scale of supersymmetry
breaking plays the role of the cutoff $\Lambda$. 

The power counting
rules, as discussed so far, are too crude to detect divergences of the
form $\ln \Lambda^2$.  Yet divergences of this form are pervasive and
extremely significant, as we shall now discuss.

\section{Running Couplings }

The problem of calculating the energy associated with a constant
magnetic field, in the more general context of an arbitrary nonabelian
gauge theory coupled to spin 0 and spin ${1\over 2}$ charged
particles, provides an excellent concrete illustration of how the
infinities of quantum field theory arise, and of how they are dealt
with.  It introduces the concept of running couplings in a natural
way, and leads directly to qualitative and quantitative results of
great significance for physics. The interactions of concern to us
appear in the Lagrangian density
\begin{equation}{\cal L} ~=~ -{1\over 4g^2} G^I_{\alpha \beta}
G^{I\alpha\beta} + \bar \psi (i\gamma^\nu D_\nu - \mu ) \psi
+\phi^\dagger(-D_\nu D^\nu - m^2 ) \phi
\label{eq:lagr}
\end{equation}
where $G^I_{\alpha \beta} \equiv
\partial_\alpha A^I_\beta - \partial_\beta A^I_\alpha -f^{IJK}
A^J_\alpha A^K_\beta $ and $D_\nu \equiv \partial_\nu + i A^I_\nu T^I$
are the standard field strengths and covariant derivative,
respectively.  Here the $f^{IJK}$ are the structure constants of the
gauge group, and the $T^I$ are the representation matrices appropriate
to the field on which the covariant derivative acts.  This Lagrangian
differs from the usual one by a rescaling $gA \rightarrow A$, which
serves to emphasize that the gauge coupling $g$ occurs only as a
prefactor in the first term. It parametrizes the energetic cost of
non-trivial gauge curvature, or in other words the stiffness of the
gauge fields.  Small $g$ corresponds to gauge fields that are
difficult to excite.  

{}From this Lagrangian itself, of course, it would
appear that the energy required to set up a magnetic field $B^I$ is
just ${1\over 2 g^2} (B^I)^2$.  This is the classical energy, but in
the quantum theory it is not the whole story.  A more accurate
calculation must take into account the effect of the imposed magnetic
field on the zero-point energy of the charged fields.  Earlier, we met
and briefly discussed a formally infinite contribution to the energy
of the ground state of a quantum field theory (specifically, the
electromagnetic field) due to the irreducible quantum fluctuations of
its modes, which mapped to an infinite number of independent harmonic
oscillators.  Insofar as only differences in energy are physically
significant, we could ignore this infinity.  But the change in the
zero-point energy as one imposes a magnetic field cannot be ignored.
It represents a genuine contribution to the physical energy of the
quantum state induced by the imposed magnetic field.  As we will soon
see, the field-dependent part of the energy also diverges.  

Postponing
momentarily the derivation, let me anticipate the form of the answer,
and discuss its interpretation.  Without loss of generality, I will
suppose that the magnetic field is aligned along a normalized,
diagonal generator of the gauge group.  This allows us to drop the
index, and to use terminology and intuition from electrodynamics
freely.  If we restrict the sum to modes whose energy is less than a
cutoff $\Lambda$, we find for the energy
\begin{equation}{\cal E}(B)
~=~ {\cal E}~ + \delta {\cal E}~= ~ {1\over 2 g^2 (\Lambda^2) } B^2 ~-
~{1\over 2} \eta B^2 ( \ln (\Lambda^2/B) + {\rm finite} )
\label{eq:eng}
\end{equation}
where 
\begin{equation}\eta ~=~ {1\over 96
\pi^2} [ - (T(R_o) -2T(R_{1\over 2}) + 2T(R_1) ) ] ~+~ {1\over 96
\pi^2} [3(-2T(R_{1\over 2}) + 8T(R_1))],
\label{eq:lncoeff}
\end{equation}
and the terms not displayed are finite
as $\Lambda \rightarrow \infty$.  The notation $g^2(\Lambda^2)$ has
been introduced for later convenience.  The factor $T(R_s)$ is the
trace of the representation for spin $s$, and basically represents the
sum of the squares of the charges for the particles of that spin.  The
denominator in the logarithm is fixed by dimensional analysis,
assuming $B >> \mu^2, m^2$.  

The most striking, and at first sight
disturbing, aspect of this calculation is that a cutoff is necessary
in order to obtain a finite result.  If we are not to introduce a new
fundamental scale, and thereby (in view of our previous discussion) endanger
locality, we must remove reference to the arbitrary cutoff $\Lambda$
in our description of physically meaningful quantities.  This is the
sort of problem addressed by the renormalization program.  Its guiding
idea is the thought that if we are working with experimental probes
characterized by energy and momentum scales well below $\Lambda$, we
should expect that our capacity to affect, or be sensitive to, the
modes of much higher energy will be quite restricted.  Thus one
expects that the cutoff $\Lambda$, which was
introduced as a calculational device to remove such modes, can be
removed (taken to infinity).  In our magnetic energy example, for
instance, we see immediately that the difference in
susceptibilities
\begin{equation}
{\cal E}(B_1) /B_1^2 - {\cal
E}(B_0)/B_0^2 ~=~ {\rm finite}
\end{equation}
is well-behaved -- that
is, independent of $\Lambda$ as $\Lambda \rightarrow \infty$.  Thus
once we measure the susceptibility, or equivalently the coupling
constant, at one reference value of $B$, the calculation gives
sensible, unambiguous predictions for all other values of $B$.  

This
simple example illustrates a much more general result, the central
result of the classic renormalization program.  It goes as follows.  A
small number of quantities, corresponding to the couplings and masses
in the original Lagrangian, that if calculated formally would diverge
or depend on the cutoff, are chosen to fit experiment.  They define
the physical, as opposed to the original, or bare, couplings.  Thus,
in our example, we can define the susceptibility to be ${1\over 2
g^2(B_0)}$ at some reference field $B_0$.  Then we have the physical or
renormalized coupling
\begin{equation}
{1\over g^2(B_0)} ~=~ {1\over
g^2(\Lambda^2) } - \eta \ln
(\Lambda^2/B_0).\label{eq:dfncouple}
\end{equation}
(In this equation I
have ignored, for simplicity in exposition, the finite terms.  These
are relatively negligible for large $B_0$.  Also, there are
corrections of higher order in $g^2$.) This of course determines the
`bare' coupling to be
\begin{equation}{1\over g^2(\Lambda^2)} ~=~
{1\over g^2(B_0)} +\eta \ln
(\Lambda^2/B_0).\label{eq:barecouple}
\end{equation}

In these terms, the central
result of diagrammatic renormalization theory is that after bare
couplings and masses are re-expressed in terms of their physical,
renormalized counterparts, the coefficients in the perturbation
expansion of any physical quantity approach finite limits, independent
of the cutoff, as the cutoff is taken to infinity.  (To be perfectly
accurate, one must also perform wave-function renormalization.  This
is no different in principle; it amounts to expressing the bare
coefficients of the kinetic terms in the Lagrangian in terms of
renormalized values.)  

The question whether this perturbation
theory converges, or is some sort of asymptotic expansion of a soundly
defined theory, is left open by the diagrammatic analysis.  This
loophole is no mere technicality, as we will soon see.  

Picking a
scale $B_0$ at which the coupling is defined is analogous to choosing
the origin of a coordinate system in geometry.  One can describe the
same physics using different choices of normalization scale, so long
as one adjusts the coupling appropriately.  We capture this idea by
introducing the concept of a running coupling defined, in accordance with
equation (\ref{eq:dfncouple}), to satisfy
\begin{equation}{d\over d \ln B} {1\over g^2(B)} ~=~
\eta.\label{eq:running} 
\end{equation}
With this definition, the choice
of a particular scale at which to define the coupling will not affect
the final result.  

It is profoundly important, however, that the
running coupling does make a real distinction between the behavior at
different mass scales, even if the original underlying theory was
formally scale invariant (as is QCD with massless quarks), and even at
mass scales much larger than the mass of any particle in the theory.
Quantum zero-point motion of the high energy modes introduces a hard
source of scale symmetry violation.  

The distinction among scales, in
a formally scale-invariant theory, embodies the phenomenon of {\it
dimensional transmutation}.  Rather than a range of theories,
parametrized by a dimensionless coupling, we have a range of theories
differing only in the value of a dimensional parameter, say (for
example) the value of $B$ at which $1/g^2(B) = 1$.  

Clearly, the
qualitative behavior of solutions of eq. (\ref{eq:running}) depends on the
sign of $\eta$.  If $\eta > 0$, the coupling $g^2(B)$ will get smaller
as $B$ grows, or in other words as we treat more and more modes as
dynamical, and approach closer to the `bare' charge.  These modes
were enhancing, or {\it antiscreening\/} the bare charge.  This is the
case of {\it asymptotic freedom}.  In the opposite case of $\eta < 0$
the coupling formally grows, and even diverges as $B$
increases. $1/g^2(B)$ goes through zero and changes sign.  On the face of
it, this would seem to indicate an instability of the theory, toward
formation of a ferromagnetic vacuum at large field strength.  This
conclusion must be taken with a big grain of salt, because when $g^2$
is large the higher-order corrections to eq. (\ref{eq:barecouple}) and
eq. (\ref{eq:running}), on which the analysis was based, cannot be
neglected.  

In asymptotically free theories, we can complete the
renormalization program in a convincing fashion.  There is no barrier
to including the effect of very large energy modes, and removing the
cutoff.  We can confidently expect, then, that the theory is
well-defined, independent of perturbation theory.  In particular,
suppose the theory has been discretized on a space-time lattice.  This
amounts to excluding the modes of high energy and momentum.  In an
asymptotically free theory one can compensate for these modes by
adjusting the coupling in a well-defined, controlled way as one shrinks
the discretization scale.  Very impressive nonperturbative
calculations in QCD, involving massive computer simulations, have
exploited this strategy.  They demonstrate the complete consistency of
the theory and its ability to account quantitatively for the masses of
hadrons.

In a non-asymptotically free theory the coupling does not become
small, there is no simple foolproof way to compensate for the missing
modes, and the existence of an underlying limiting theory becomes
doubtful.

Now
let us discuss how $\eta$ can be calculated.  The two terms in
eq. (\ref{eq:lncoeff}) correspond to two distinct physical effects.  The
first is the convective, diamagnetic (screening) term.  The overall
constant is a little tricky to calculate, and I do not have space to
do it here.  Its general form, however, is transparent.  The effect is
independent of spin, and so it simply counts the number of components
(one for scalar particles, two for spin-1/2 or massless spin-1, both
with two helicities).  It is screening for bosons, while for fermions
there is a sign flip, because the zero-point energy is negative for
fermionic oscillators.  

The second is the paramagetic spin
susceptibility. For a massless particle with spin $s$ and gyromagnetic
ratio $g_m$ the energies shift, giving rise to the altered zero-point
energy
\begin{equation}
\Delta {\cal E} ~=~ \int^{E=\Lambda}_0
{d^3k\over (2\pi)^3} {1\over 2} (\sqrt {k^2 + g_m s B} + \sqrt{k^2 -
g_msB} - 2 \sqrt {k^2}).\label{eq:eshift}
\end{equation}
This is readily
calculated as
\begin{equation}
\Delta {\cal E} ~=~ - B^2 (g_ms)^2
{1\over 32 \pi^2} \ln ({\Lambda^2\over B}).
\label{eq:enum}
\end{equation}
With $g_m = 2$, $s=1$ (and $T=1$) this
is the spin-1 contribution, and with $g_m =2$, $s = {1\over 2}$, after
a sign flip, it is the spin-${1\over 2}$ contribution.  The preferred
moment $g_m = 2$ is a direct consequence of the Yang-Mills and Dirac
equations, respectively. 

This elementary calculation gives us a nice
heuristic understanding of the unusual antiscreening behavior of
nonabelian gauge theories.  It is due to the large paramagnetic
response of charged vector fields.  Because we are interested in very
high energy modes, the usual intuition that charge will be screened,
which is based on the electric response of heavy particles, does not
apply.  Magnetic interactions, which can be attractive for like
charges (paramagnetism) are, for highly relativistic particles, in no
way suppressed.  Indeed, they are numerically dominant.  

Though I
have presented it in the very specific context of vacuum magnetic
susceptibility, the concept of running coupling is much more widely
applicable.  The basic heuristic idea is that in analyzing processes
whose characteristic energy-momentum scale (squared) is $Q^2$, it is
appropriate to use the running coupling at $Q^2$, i.e. in our earlier
notation $g^2(B = Q^2)$.  For in this way we capture the dynamical
effect of the virtual oscillators which can be appreciably excited,
while avoiding the formal divergence encountered if we tried to
include all of them (up to infinite mass scale).  At a more formal
level, use of the appropriate effective coupling allows us to avoid
large logarithms in the calculation of Feynman graphs, by normalizing
the vertices close to where they need to be evaluated.  There is a
highly developed, elaborate chapter of quantum field theory which
justifies and refines this rough idea into a form where it makes
detailed, quantitative predictions for concrete experiments.  I will
not be able to do proper justice to the difficult, often heroic labor
that has been invested, on both the theoretical and the experimental
sides, to yield Figure 1; but it is appropriate to remark that quantum
field theory gets a real workout, as calculations of two- and even
three-loop graphs with complicated interactions among the virtual
particles are needed to do justice to the attainable experimental
accuracy.  

\begin{figure}
\centerline{\psfig{figure=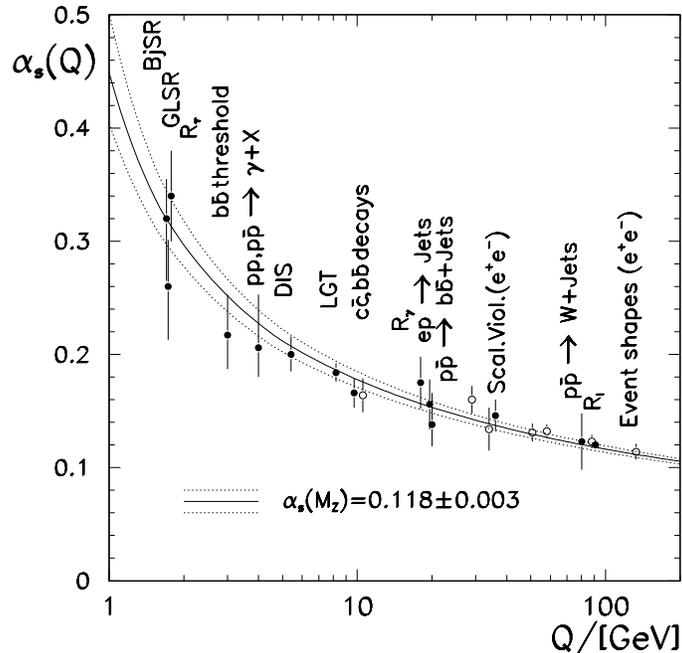,width=3.5in}}
\caption{Comparison of theory and experiment in QCD,
illustrating the running of couplings.  Several of the points on this
curve represent hundreds of independent measurements, any one of which
might have falsified the theory. Figure from M. Schmelling, hep-ex/9701002. }
\label{fig1}
\end{figure}

An interesting feature visible in
Figure 1 is that the theoretical prediction for the coupling focuses
at large $Q^2$, in the sense that a wide range of values at small
$Q^2$ converge to a much narrower range at larger $Q^2$.  Thus even
crude estimates of what are the appropriate scales (e.g., one expects
$g^2(Q^2)/4 \pi \sim 1$ where the strong interaction is strong, say
for 100 Mev $\lsim \sqrt{Q^2} \lsim$ 1 Gev) allow one to predict the value of
$g^2(M_Z^2)$ with $\sim$10\% accuracy.  The original idea of Pauli and
others that calculating the fine structure constant was the next great
item on the agenda of theoretical physics now seems misguided.  We see
this constant as just another running coupling, neither more nor less
fundamental than many other parameters, and not likely to be the most
accessible theoretically.  But our essentially parameter-free approximate
determination of the observable strong interaction analogue of the
fine structure constant realizes a form of their dream. 

The
electroweak interactions start with much smaller couplings at low mass
scales, so the effects of their running are less dramatic (though they
have been observed).  Far more spectacular than the modest quantitative
effects we can test directly, however, is the conceptual breakthrough
that results from 
application of these ideas to unified models of the strong,
electromagnetic, and weak interactions.  

The different components of
the standard model have a similar mathematical structure, all being
gauge theories.  Their common structure encourages the speculation
that they are different facets of a more encompassing gauge symmetry,
in which the different strong and weak color charges, as well as
electromagnetic charge, would all appear on the same footing.  The
multiplet structure of the quarks and leptons in the standard model
fits beautifully into small representations of unification groups such
as $SU(5)$ or $SO(10)$.  There is the apparent difficulty, however,
that the coupling strengths of the different standard model
interactions are widely different, whereas the symmetry required for
unification requires that they share a common value.The running of
couplings suggests an escape from this impasse.  Since the strong,
weak, and electromagnetic couplings run at different rates, their
inequality at currently accessible scales need not reflect the
ultimate state of affairs.  We can imagine that spontaneous symmetry
breaking -- a soft effect -- has hidden the full symmetry of the
unified interaction.  What is really required is that the fundamental,
bare couplings be equal, or in more prosaic terms, that the running
couplings of the different interactions should become equal beyond
some large scale.  

Using simple generalizations of the formulas
derived and tested in QCD, we can calculate the running of couplings,
to see whether this requirement is satisfied in reality.  
In doing so one must make
some hypothesis about the spectrum of virtual particles.  If there are
additional massive particles (or, better, fields) that have not yet
been observed, they will contribute significantly to the running of
couplings once the scale exceeds their mass. Let us first consider the
default assumption, that there are no new fields beyond those that
occur in the standard model.  The results of this calculation are
displayed in Figure 2. 

\begin{figure}
\centerline{\psfig{figure=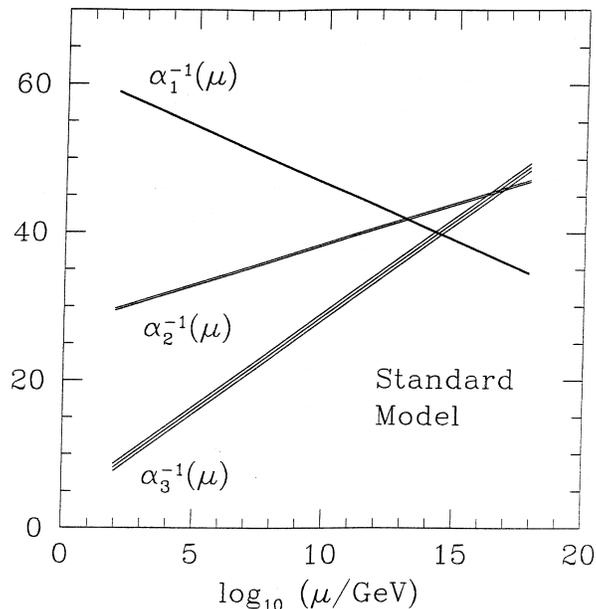,width=3.25in}}
\caption{Running of the couplings
extrapolated toward very high scales, using just the fields of the
standard model.  The couplings do not quite meet.  Experimental
uncertainties in the extrapolation are indicated by the width of the
lines. Figure courtesy of K. Dienes.}
\label{fig2}
\end{figure}

Considering the enormity of the extrapolation this calculation works 
remarkably well, but the accurate experimental data indicates
unequivocally that something is wrong.  There is one particularly
attractive way to extend the standard model, by including
supersymmetry.  Supersymmetry cannot be exact, but if it is only
mildly broken (so that the superpartners have masses $\lsim$ 1 Tev) it
can help explain why radiative corrections to the Higgs mass
parameter, and thus to the scale of weak symmetry breaking, are not
enormously large.  In the absence of supersymmetry power counting
would indicate a hard, quadratic dependence of this parameter on the
cutoff.  Supersymmetry removes the most divergent contribution, by
cancelling boson against fermion loops.  If the masses of the
superpartners are not too heavy, the residual finite contributions due
to supersymmetry breaking will not be too large.  

The minimal
supersymmetric extension of the standard model, then, 
makes semi-quantitative
predictions for the spectrum of virtual particles starting at 1 Tev or
so.  Since the running of couplings is logarithmic, it is not
extremely sensitive to the unknown details of the supersymmetric mass
spectrum, and we can assess the impact of supersymmetry on the
unification hypothesis quantitatively.  The results, as shown in
Figure 3, are quite encouraging.

\begin{figure}
\centerline{\psfig{figure=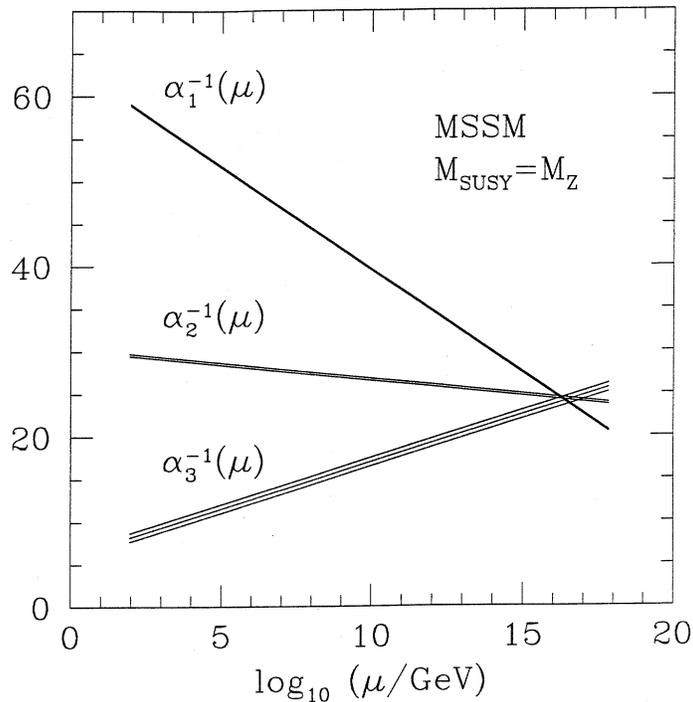,width=3.5in}}
\smallskip
\caption{ Running of the
couplings extrapolated to high scales, including the effects of
supersymmetric particles starting at 1 Tev.  Within experimental and
theoretical uncertainties, the couplings do meet. Figure courtesy of
K. Dienes. }
\label{fig3}
\end{figure}

With all its
attractions, there is one general feature of supersymmetry that is
especially challenging, and deserves mention here.  We remarked
earlier how the standard model, without supersymmetry, features a
near-perfect match between the generic symmetries of its renormalizable
interactions and the observed symmetries of the world.  With
supersymmetry, this feature is spoiled.  The scalar superpartners of
fermions are represented by fields of mass dimension one.  This means
that there are many more possibilities for low dimension (including
renormalizable) interactions that violate flavor symmetries including
lepton and baryon number.  It seems that some additional principles,
or special discrete symmetries, are required in order to suppress
these interactions sufficiently.

A notable result of the unification of couplings calculation,
especially in its supersymmetric form, is that the unification occurs
at an energy scale which is enormously large by the standards of
traditional particle physics, perhaps approaching $10^{16-17}$ Gev.
{}From a phenomenological viewpoint, this is fortunate.  The most
compelling unification schemes merge quarks, antiquarks, leptons, and
antileptons into common multiplets, and have gauge bosons mediating
transitions among all these particle types.  Baryon number violating
processes almost inevitably result, whose rate is inversely
proportional to the fourth power of the gauge boson masses, and thus
to the fourth power of the unification scale.  Only for such large
values of the scale is one safe from experimental limits on nucleon
instability.  From a theoretical point of view the large scale is
fascinating because it brings us from the internal logic of the
experimentally grounded domain of particle physics to the threshold of
quantum gravity, as we shall now discuss.



\section{Limitations?}

So much for the successes, achieved and anticipated, of quantum field
theory.  The fundamental limitations of quantum field theory, if any,
are less clear.  Its application to gravity has certainly, to date,
been much less fruitful than its triumphant application to describe
the other fundamental interactions.   

All existing experimental
results on gravitation are adequately described by a very beautiful,
conceptually simple classical field  theory -- Einstein's general
relativity.  It is easy to incorporate this  theory into our
description of the  world based on quantum field theory, by allowing a
minimal coupling to the fields of the standard model -- that is, by
changing ordinary into covariant derivatives, multiplying with
appropriate factors of $\sqrt g$, and adding an Einstein-Hilbert
curvature term.   
The resulting theory --  with the convention that we
simply ignore quantum corrections involving virtual gravitons -- is
the foundation of our working description of the physical world.  As a
practical matter, it works very well indeed.    

Philosophically, however, it
might be disappointing if it were too straightforward to construct a
quantum theory of gravity.  One of the great visions of natural
philosophy, going back to Pythagoras, is that the properties of the
world are determined uniquely by mathematical principles.  A modern
version of this vision was formulated by Planck, shortly after he
introduced his quantum of action.  By appropriately combining the
physical constants $c$, $\hbar$ as units of velocity and action,
respectively, and the Planck mass $M_{\rm Planck} = \sqrt {\hbar c
\over G}$ as the unit of mass, one can construct any unit of
measurement used in physics.   
Thus the unit of energy is $M_{\rm Planck} c^2$, the unit of electric
charge is $\sqrt {\hbar c}$, and so forth.  On the other hand, one
cannot form a pure number from these three physical constants.  Thus
one might hope that in a physical theory where $\hbar$, $c$, and $G$
were all profoundly incorporated, all physical quantities could be
expressed in natural units as pure numbers.

Within its domain, QCD achieves  something very close to
this vision --  actually, in a {\it more\/} ambitious form!   Indeed,
let us idealize the world of the strong interaction slightly, by
imagining that there were just two quark species with vanishing
masses.   Then from the two integers 3 (colors) and 2 (flavors),
$\hbar$, and $c$ -- with {\it no\/} explicit mass parameter -- a
spectrum of hadrons, with mass ratios and other properties close
to those observed in reality, emerges by calculation.   
The overall unit of mass is indeterminate, but this ambiguity has no
significance within the theory itself.  

The ideal
Pythagorean/Planckian theory would not contain any pure numbers as
parameters.  (Pythagoras might have excused a few small integers).
Thus, for example, the value $m_e/M_{\rm Planck} \sim 10^{-22}$ of the
electron mass in Planck units would emerge from a dynamical
calculation.
This ideal might be overly ambitious, yet it seems reasonable to
hope that significant constraints among physical observables will
emerge from the inner requirements of a quantum theory which
consistently incorporates gravity.  Indeed, as we have already seen, one does
find significant constraints among the parameters of the standard
model by requiring that the strong, weak, and electromagnetic
interactions emerge from a unified gauge symmetry; so there is
precedent for results of this kind.

The unification of couplings calculation provides not only an
inspiring model, 
but also direct
encouragement for the Planck program, in two important respects.  First,
it points to a symmetry breaking scale remarkably close to the Planck
scale (though apparently smaller by $10^{-2}-10^{-3}$), so there are
pure numbers with much more `reasonable' values than $10^{-22}$ to
shoot for.  Second, it shows quite concretely how very large scale
factors can be controlled by modest ratios of coupling strength, due
to the logarithmic nature of the running of couplings -- so that
$10^{-22}$ may not be so `unreasonable' after all.  

Perhaps it is fortunate, then, that the straightforward, minimal
implementation of general relativity as a quantum field theory --
which lacks the desired constraints -- runs into problems.  The
problems are of two quite distinct kinds.  First, the renormalization
program fails, at the level of power-counting.  The Einstein-Hilbert
term in the action comes with a large prefactor $1/G$, reflecting the
difficulty of curving space-time.  If we expand the Einstein-Hilbert
action around flat space in the form
\begin{equation}
g_{\alpha \beta} =
\eta_{\alpha\beta} +  \sqrt G h_{\alpha\beta}
\end{equation}
we find
that  the quadratic terms give a properly normalized spin-2  graviton
field $h_{\alpha\beta}$ of mass dimension 1, as the powers of $G$
cancel.   But the higher-order terms, which represent interactions,
will be accompanied by positive powers of $G$.  Since $G$ itself has
mass dimension -2, these are non-renormalizable interactions.
Similarly for the couplings of gravitons to matter.  Thus we can expect
that ever-increasing powers of $\Lambda/M_{\rm Planck}$ will appear in
multiple virtual graviton exchange, and it will be impossible to
remove the cutoff.  

Second, one of the main qualitative features of
gravity -- the weightlessness of empty space, or the vanishing of the
cosmological constant -- is left unexplained.   Earlier we mentioned
the divergent zero-point energy characteristic of generic quantum
field theories.  For  purposes of non-gravitational physics only energy
differences are meaningful, and we can sweep this problem under the
rug.  But gravity ought  to see this energy.   Our perplexity
intensifies  when we recall that according to the standard model, and
even more so in its unified extensions, what we commonly regard as
empty space is full of condensates, which again one would expect to
weigh far more than observation allows.  The failure, so far, of
quantum field theory to meet these challenges might reflect a basic
failure of principle, or merely that the appropriate symmetry
principles and degrees of freedom, in terms of which the theory should
be formulated, have not yet been identified.   

Promising insights toward construction of a quantum theory including
gravity are coming from investigations in string/M theory, as
discussed elsewhere in this volume.  Whether these investigations will
converge toward an accurate description of nature, and if so whether
this description will take the form of a local field theory (perhaps
formulated in many dimensions, and including many fields beyond those
of the standard model) , are questions not yet decided.  It is
interesting, in this regard, briefly to consider the rocky
intellectual history of quantum field theory.

After the initial successes of the 1930s, already mentioned above,
came a long period of disillusionment.  Initial attempts to deal with
the infinities that arose in calculations of loop graphs in
electrodynamics, or in radiative corrections to beta decay, led only
to confusion and failure.  Similar infinities plagued Yukawa's pion
theory, and it had the additional difficulty that the coupling
required to fit experiment is large, so that tree graphs provide a
manifestly poor approximation.  Many of the founders of quantum
theory, including Bohr, Heisenberg, Pauli, and (for different reasons)
Einstein and Schrodinger, felt that further progress required a
radically new innovation.  This innovation would be a revolution of
the order of quantum mechanics itself, and would introduce a new
fundamental length.

Quantum electrodynamics was resurrected in the late 1940s, largely
stimulated by developments in experimental technique.  These
experimental developments made it possible to study atomic processes
with such great precision, that the approximation afforded by keeping
tree graphs alone could not do them justice.  Methods to extract
sensible finite answers to physical questions from the jumbled
divergences were developed, and spectacular agreement with experiment
was found -- all without changing electrodynamics itself, or departing
from the principles of relativistic quantum field theory.  

After this
wave of success came another long period of disillusionment.  The
renormalization methods developed for electrodynamics did not seem to
work for weak interaction theory.  They did suffice to define a
perturbative expansion of Yukawa's pion theory, but the strong
coupling made that limited success academic (and it came to seem
utterly implausible that Yukawa's schematic theory could do justice to
the wealth of newly discovered phenomena).  In any case, as a
practical matter, throughout the 1950s and 1960s a flood of
experimental discoveries, including new classes of weak processes and
a rich spectrum of hadronic resonances with complicated interactions,
had to be absorbed and correlated.  During this process of pattern
recognition the elementary parts of quantum field theory were used
extensively, as a framework, but deeper questions were put off.  Many
theorists came to feel that quantum field theory, in its deeper
aspects, was simply wrong, and would need to be replaced by some
S-matrix or bootstrap theory; perhaps most thought it was irrelevant,
or that its use was premature, especially for the strong interaction.

As it became clear, through phenomenological work, that the weak
interaction is governed by current$\times$current interactions with
universal strength, the possibility to ascribe it to exchange of
vector gauge bosons became quite attractive.  Models incorporating the
idea of spontaneous symmetry breaking to give mass to the weak gauge
bosons were constructed.  It was conjectured, and later proved, that
the high degree of symmetry in these theories allows one to isolate
and control the infinities of perturbation theory.  One can carry out
a renormalization program similar in spirit, though considerably more
complex in detail, to that of QED.  It is crucial, here, that
spontaneous symmetry breaking is a very soft operation.  It does not
significantly affect the symmetry of the theory at large momenta,
where the potential divergences must be cancelled.  

Phenomenological work on the strong interaction made it increasingly
plausible that the observed strongly interacting particles -- mesons
and baryons -- are composites of more basic objects.  The evidence was
of two disparate kinds: on the one hand, it was possible in this way
to make crude but effective models for the observed spectrum with
mesons as quark-antiquark, and baryons as quark-quark-quark, bound
states; and on the other hand, experiments provided evidence for hard
interactions of photons with hadrons, as would be expected if the
components of hadrons were described by local fields.  The search for
a quantum field theory with appropriate properties led to a unique
candidate, which contained both objects that could be identified with
quarks and an essentially new ingredient, color gluons.

These quantum field theories of the
weak and strong interactions were dramatically confirmed by subsequent
experiments, and have survived exceedingly rigorous testing over the
past two decades.  They make up the Standard Model.  During this
period the limitations, as well as the very considerable virtues, of
the Standard Model have become evident.  Whether the next big step
will require a sharp break from the principles of quantum field theory
or, like the previous ones, a better appreciation of its
potentialities, remains to be seen.

\section*{Acknowledgments}
I wish to thank S. Treiman for extremely helpful guidance, and
M. Alford, K. Babu, C. Kolda, and J. March-Russell for reviewing the
manuscript.  F.W. is supported in part by DOE grant DE-FG02-90ER40542

\section*{References}
For further information about quantum field theory, the
reader may wish to consult:

\begin{enumerate}

\item T.P. Cheng and L.F. Li, Gauge Theory of Elementary Particle
Physics, (Oxford, 1984).

\item M. Peskin and D. Schroeder,  Introduction to Quantum Field
Theory, (Addison-Wesley, 1995).

\item S. Weinberg,  The Quantum Theory of Fields, I, (Cambridge,
1995) and  The Quantum Theory of Fields, II, (Cambridge, 1996).

\end{enumerate}

\end{document}